\documentclass[prc,showpacs,twocolumn]{revtex4}
\usepackage{graphicx}
\begin{document}

\title{
Proton enhancement at large $p_T$ at LHC without structure in associated-particle
distribution}

\author{Rudolph C. Hwa$^1$ and  C.\ B.\ Yang$^{1,2}$}
\affiliation
      {$^1$Institute of Theoretical Science and Department of
Physics\\ University of Oregon, Eugene, OR 97403-5203, USA}
\affiliation{
$^2$Institute of Particle Physics, Hua-Zhong Normal University,
Wuhan 430079, P.\ R.\ China}

\begin{abstract}
The production of pions and protons in the $p_T$ range between 10 and 20 GeV/c for
Pb+Pb collisions at LHC is studied in the recombination model. It is shown that the
dominant mechanism for hadronization is the recombination of shower partons from
neighboring jets when the jet density is high. Protons are more copiously produced
than pions in that $p_T$ range because the coalescing partons can have lower momentum
fractions, but no thermal partons are involved. The proton-to-pion ratio can be as high as 20. When such high $p_T$ hadrons are used as
trigger particles, there will not be any associated particles that are not in the
background.
\end{abstract}

\maketitle
Particle production at large $p_T$ in $pp$ collisions is well understood in
terms of the fragmentation of partons after hard scattering, so the ratio
of proton to pion can be related to the ratio of the corresponding fragmentation
functions (FF), $D^p/D^{\pi}$, which is roughly of the order of $10^{-1}$.
In central heavy-ion collisions at RHIC the $p/\pi$ ratio, $R_{p/\pi}$, has been found
to be greater than 1 at $p_T \sim 3$ GeV/c before diminishing to $\sim 0.3$ at higher
$p_T$ \cite{ssa,jd}.  Clearly, an alternative mechanism of hadronization is at play at intermediate $p_T$  \cite{hy,gr,fr}.  The question we ask here is whether the same phenomenon
will occur in heavy-ion collisions at LHC.  In this paper we give reasons to expect a new phenomenon to take place that
will make $R_{p/\pi}$ to be even larger in the  $10 < p_T < 20$ GeV/c range. We further predict that there are no peaks in the
$\Delta\phi$ distribution of the particles associated with a trigger in that $p_T$ range.

The alternative mechanism of hadronization referred to above is parton recombination.
The dominance of recombination over fragmentation at high $p_T$ (now regarded as
intermediate) was recognized long before there was any heavy-ion data on the subject
\cite{rch}.  The high-quality data now available from RHIC  make possible
more detailed treatment of the recombination process where thermal and shower partons
can both contribute at intermediate $p_T$ \cite{hy2}.  The importance of thermal
partons recedes as $p_T$ is increased to above 10 GeV/c.  At such high $p_T$ at
RHIC the dominant process is fragmentation. However, at LHC where the density of
jets is very high, a new phenomenon arises where the recombination of shower partons
in neighboring jets can make a significant contribution.  We investigate here the $p_T$
distributions of pion and proton in the range $10 < p_T < 20$ GeV/c and show that the
$p/\pi$ ratio can be very large, which can therefore serve as an excellent
signature of the hitherto unrealized hadronization process.

The study of hadron production in central heavy-ion collisions at RHIC in the
intermediate $p_T$ range has been characterized by three features in \cite{hy2}:
(a) suppression due to energy loss in the medium is represented by a factor $\xi$;
(b) jets produced near the surface create shower partons whose distributions have
independently been determined \cite{hy3}; (c) medium effect on hadron production
is taken into account by  the recombination of thermal and shower partons.
The thermal parton distribution is determined by fitting the low-$p_T$ spectra
( $< 2$ GeV/c) in the recombination model.  As we look ahead toward LHC, we do not
know what the suppression factor $\xi$ is, nor do we have the soft spectra to provide
information about the thermal partons.  What we do know is the shower parton
distributions (SPD), which were determined from the  FFs
independent of the colliding system.  Thus we want to make predictions that rely
heavily on the SPD, and minimally on the thermal partons.  To that end we consider
$p_T$ range greater than 10 GeV/c in order to minimize the contribution from the
thermal partons.  On the other hand, we do not want $p_T$ to be too high, since the
central theme of this investigation is the recombination of shower partons from
neighboring jets, for which the density of jets must be high.  Thus we focus our
attention in the region $10 < p_T < 20$ GeV/c, which is relatively narrow in the
overall range available at LHC, reaching up to 100 GeV/c.  For $\xi$ we shall just
consider two typical values that seem reasonable.

Our formulation of parton recombination is one-dimensional in the direction of the
detected hadron.  If there is only one jet, then the recombination of two shower
partons in the same jet is equivalent to fragmentation into a meson, since that
is how shower partons are defined in the first place \cite{hy3}.  The corresponding
invariant distribution for a hadron $h$  is
\begin{eqnarray}
 p{dN^{(1)}_h  \over  dp} = \xi  \sum_i  \int dk k f_i(k)  {p \over  k} D^h_i
 \left( p \over  k\right)  ,
 \label{1}
\end{eqnarray}
where $p$ is an abbreviation for $p_T$  and
$k$ is the transverse momentum of the scattered hard parton $i$; its distribution
in a heavy-ion collision is $f_i(k)$ that  takes into account the parton distribution
functions, shadowing effect in the nuclei, hard scattering cross section, etc., as
parameterized in Ref.\ \cite{sr}.  $D^h_i (z)$ is the FF for a parton $i$ to a hadron
$h$; $\xi$ is the suppression factor that was found to be $\xi = 0.07$ for RHIC
\cite{hy2}.  We shall use $\xi = 0.01$ and $0.03$ as representative values for LHC
in the following.

The new component that we consider here is the two-jet contribution $pdN^{(2)}_h/dp$,
which is given by
 \begin{eqnarray}
 p{dN^{(2)}_h  \over  dp} = \xi^2  \sum_{i,i'}  \int dk dk' k f_i(k) k' f_{i'}(k')
 \Gamma (s, k, k' ) \nonumber\\
\times\int(\prod_\ell {dp_\ell\over p_\ell}) F_{ii'}(k,k'; p_1, p_2, [p_3]) R_h (p_1, p_2, [p_3], p)  ,
 \label{2}
\end{eqnarray}
where $F_{ii'}$ denotes the SPDs in the two jets and $R_h$ the recombination function
(RF) for the formation of $h$ at $p$.  The symbol $[p_3]$ means that it is present
(absent) for baryon (meson) production.  $\Gamma (s, k, k')$
is the overlap function between the two jets, which depends not only on the c.m.~energy
$s$, but also on the hard parton momentum vectors $\vec{k}$ and $\vec{k'}$, and the
widths of their jet cones.  That is the main quantity that we do not have sufficient
information on for collisions at LHC.  We shall approximate it by an average quantity
$\Gamma$ and pull it out of the integral, using several possible values for it.
The other parts of the integrand are calculable.

For pion production we have
\begin{equation}
F_{ii'}(k,k'; p_1, p_2)  = S^j_i \left({p_1 \over  k}\right)  S^{j'}_{i'} \left({p_2
\over  k'}\right) ,
 \label{3}
\end{equation}
where $S^j_i(z)$ is the SPD of quark $j$ with momentum fraction $z$ in a jet initiated
by parton $i$   \cite{hy3}.  The difference between (\ref{3})
and the corresponding equation in the one-jet case is that $p_1$ and $p_2$ are not bounded
by $p_1 + p_2 \leq k$, the only jet momentum.  Here, $p_1$ and $p_2$ can both be large
(though each bounded by $k$ and $k'$, respectively), resulting in a large pion momentum
at $p = p_1 + p_2$.  For proton production we have
\begin{eqnarray}
F_{ii'}(k,k';p_1,p_2,p_3)=S^j_i\left({p_1 \over  k}\right)\left\{S^{j'}_{i'}\left({p_2 \over  k'}\right),S^{j''}_{i'} \left({p_3 \over  k'-p_2}\right)\right\},
 \label{4}
\end{eqnarray}
where the curly brackets imply a symmetrization of $p_2$ and $p_3$.  Since there exists
symmetry in $i$ and $i'$, there is no need for us to consider two shower partons in
jet $i$.  The partons $j$, $j'$ and $j''$ are to be permuted among the types $u$,
$u$ and $d$ for proton formation, and are not indicated explicitly in Eq.\ (\ref{2}).
The partons $i$ and $i'$ are to be summed over all species $u$, $d$,  $s$, $\bar{u}$,
$\bar{d}$,  $\bar{s}$ and $g$.  Thus there are many terms in the summation that involves
valence, sea and gluon SPDs.   The RFs $R_h$  have been given before in Ref. \cite{hy2} and in references cited therein. They all have the momentum constraints $\delta(\sum_i p_i-p)$.

The SPDs $S^j_i(z)$ have been parameterized in \cite{hy3} as
\begin{eqnarray}
S^j_i(z) = Az^a (1 - z)^b (1+cz^d).
\label{5}
\end{eqnarray}
For numerical estimates let us focus on the shower partons generated by a gluon, since
for Pb+Pb collisions at LHC the dominant hard partons at high $k$ are gluons \cite{sr}.
For $i = g$ and $j = q$ ($u$ or $d$), the parameters in Eq. (\ref{5}) are:  $A = 0.811$,
$a = -0.056$, $b = 2.547$, $c = -0.176$, and $d = 1.2$.  Thus we have $S^q_g (1/3) = 0.3$,
while $S^q_g (1/2) = 0.13$.  There are therefore more than twice the number of light
quarks in the gluon jet at $z = 1/3$ than at $z = 1/2$.  That huge difference leads to higher probability for proton formation than for pion.

To calculate the $p_T$ distributions of $\pi$ and $p$, we need to sum over all hard
partons $i$ and $i'$, and integrate over all their momenta $k$ and $k'$.  Let us write \begin{eqnarray}
{dN_h \over p_T dp_T}  = H^{(1)}_h (p_T, \xi) + \Gamma H^{(2)}_h (p_T, \xi) ,
\label{6}
\end{eqnarray}
where $H^{(1)}_h (p_T, \xi)$ is the one-jet contribution given in Eq.\ (\ref{1}),
multiplied by $p^{-2}_T$, and $H^{(2)}_h (p_T, \xi)$ is the two-jet integral in
Eq.\ (\ref{2}), also multiplied by $p^{-2}_T$, but without $\Gamma (s, k, k')$,
which is approximated by the constant overlap factor $\Gamma$ taken outside the
integral for $p_T$ in the range $10 < p_T < 20$ GeV/c.  Thus the total $p_T$
distribution that we can calculate depends on two parameters $\xi$ and $\Gamma$,
which depend in turn on the suppression of hard partons in their emergence from
the dense medium, their density outside the medium, and the overlap of showers
from neighboring jets.  To account for all possibilities, we allow $\Gamma$ to
vary over a wide range, i.e., $\Gamma = 10 ^{-n}$ where $n = 1, 2, 3$ and $4$.

\begin{figure}[tbph]
\includegraphics[width=0.4\textwidth,height=0.5\textwidth]{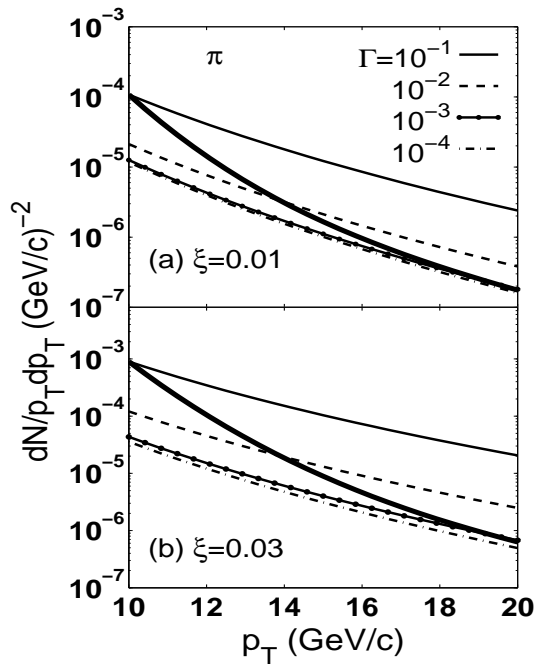}
\caption{Transverse momentum distributions of pions for two values of the nuclear
suppression factor $\xi$ and for various fixed values of the probability $\Gamma$ of overlap of neighboring jets. The heavy solid line represents the distribution when $\Gamma(p_T)$ is taken to decrease as $p_T^{-7}$.}
\end{figure}

In Fig.\ 1 we show the results of our calculation of pion production for (a)
$\xi = 0.01$ and (b) $\xi = 0.03$.  The four curves in each panel are for the
four values of $\Gamma$ indicated in the legend.  As $\Gamma$ becomes infinitesimal,
i.e.\ no overlap between two neighboring jet cones, the curves approach limiting
curves (not shown) that correspond to the fragmentation from one jet.  Thus what
stand above the limiting curves are all due to the recombination of the shower
partons from two different jets, and they can be quite large.  The reason is
that in recombination the pion momentum $p$ is $p_1 + p_2$ and is not limited
by either $k$ or $k'$ in  Eq.\ (\ref{3}), as in the case of one-jet fragmentation.
Put differently, for any fixed value of $p$, the hard parton momenta $k$ and $k'$ in
the two-jet case can be lower than the only parton momentum $k$ in the one-jet case,
so the rate of production can be higher.

The approximation of constant $\Gamma$ can be removed by letting $\Gamma$ vary with $p_T$ in accordance to a power law, since the jet production probability decreases as $p_T^{-n}$. If we start with $\Gamma=10^{-1}$ at $p_T=10$ GeV/c, which is reasonable for an expected jet multiplicity of 100 in $|\eta|<0.5$, and if we let it decrease as $\Gamma(p_T)\propto p_T^{-7}$, then we get the pion distribution as shown by the heavy solid lines in Fig.\ 1. Evidently, the contribution from two-jet recombination becomes insignificant at $p_T\sim 20$ GeV/c.

\begin{figure}[tbph]
\includegraphics[width=0.4\textwidth,height=0.5\textwidth]{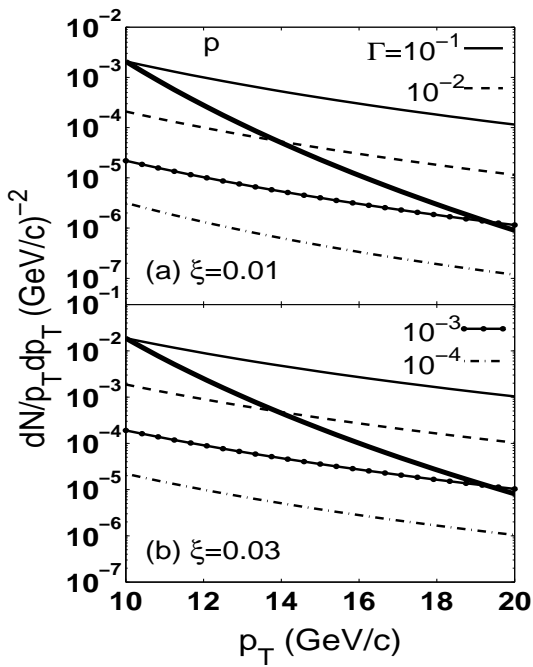}
\caption{Same as in Fig.\ 1 but for proton.}
\end{figure}

The $\Gamma$ dependence of the proton distribution is quite different, as shown
in Fig.\ 2.  At small $\Gamma$ there is no hint of the curves approaching limiting curves.
That is, the fragmentation of a parton into a proton has very low probability.  However,
the contribution from two jets grows very fast with increasing $\Gamma$.  For $\Gamma =
10^{-1}$ the rate of proton production is more than an order of magnitude higher than that
of pion.  The reason is similar to that given for pion, but the effect is more amplified.
For a fixed $p = p_1 + p_2 + p_3$ the hard parton momenta $k$ or $k'$ need not be greater
than $p$ as required by fragmentation.  Furthermore, with three shower partons
contributing to $p$ the momentum fractions of each can be lower, resulting in higher
yield.

When we allow $\Gamma(p_T)$ to decrease with $p_T$ as in the case of pion considered above, the proton distribution is shown by the heavy solid line in Fig.\ 2. It now decreases far more rapidly with $p_T$ , but is still higher than the 1-jet contribution (not shown) even at $p_T=20$ GeV/c.

Knowing the $p_T$ distributions of pion and proton, we can take their ratio and exhibit
the remarkable behavior of the $p$ to $\pi$ ratio, $R_{p/\pi}$, as shown in Fig.\ 3.  First of all, for constant $\Gamma$ the ratio increases with $p_T$, showing the dominance of three-quark recombination for proton formation. However, when the decrease of $\Gamma(p_T)$ with $p_T$ is taken into account, $R_{p/\pi}$ shows also a decrease with $p_T$ (in heavy solid lines) rather than increases in the constant $\Gamma$ cases. The range of the ratio decreasing from 20 to 5 is significantly higher than that of the 1-jet case, as shown by the light solid lines in Fig.\ 3. A difference of 2 orders of magnitude  is a spectacular manifestation of a new hadronization process at high $p_T$.
 Figure 3 differs from the prediction of Ref.\ \cite{fm} because only
the fragmentation term corresponding to $\Gamma=0$ is shown there.

\begin{figure}[tbph]
\includegraphics[width=0.45\textwidth,height=0.5\textwidth]{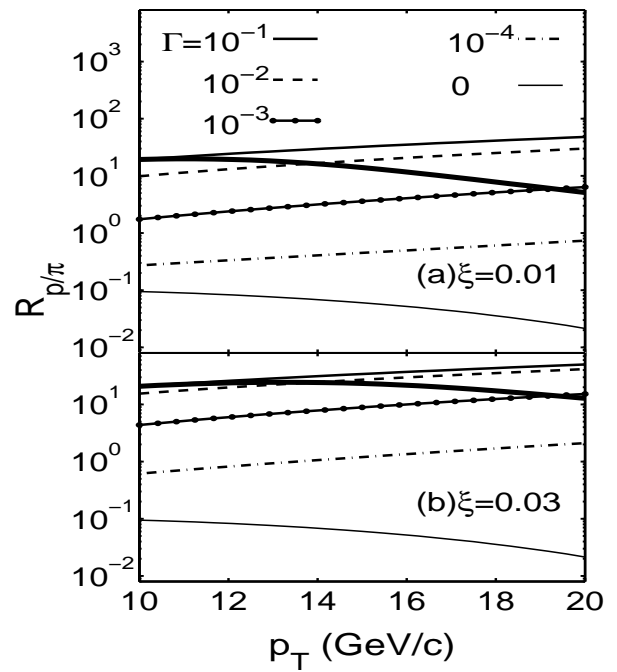}
\caption{The proton-to-pion ratio of the $p_T$ distributions at LHC. The heavy solid lines show the ratio when $\Gamma(p_T)$ is taken to decrease as $p_T^{-7}$. The light solid lines present  the ratio when  only a single jet contributes.}
\end{figure}

For $p_T>20$ GeV/c the mean jet multiplicity is less than 2 in $|\eta|<0.5$, and $\Gamma(p_T)$ is expected to decrease faster than $p_T^{-7}$; thus the heavy solid lines should rapidly join the light solid lines at higher $p_T$.  The pion and proton distributions are roughly proportional to $\xi^2$, so the ratio is approximately independent of $\xi$. Furthermore, since $\Gamma(p_T)$ is the same for the two distributions, $R_{p/\pi}$ does not depend crucially on what $\Gamma$ is exactly at $p_T=10$ GeV/c.

In addition to the surprisingly large $p/\pi$ ratio, there is another feature about
hadron production at large $p_T$ at LHC that is unusual compared to what is routine
at RHIC. When there is a large-$p_T$ particle detected at RHIC, it can be used to
serve as a trigger, and the associated particles are shown to have structure with
peaks and valleys in $\Delta\eta$ and $\Delta\phi$ distributions \cite{jd,ja,ssa1,jj}.
To find the peaks in the associated particle distribution at RHIC it is necessary to
make background subtraction so that the jet effects can be clearly isolated from the
statistical distribution of particles from the bulk medium that is always present with
or without jets. It should be noted that to have a high-$p_T$ particle at RHIC is a
rare event, and that the jet phenomenon is a signal of hard scattering that stands out
above the background.

At LHC, on the other hand, jets are not rare; indeed, we have used the high density
of jets to calculate the contribution of two jets to the formation of a single hadron.
If jets are numerous in every event, then they are part of the background. Furthermore,
for the hadrons that are detected in the $10<p_T<20$ GeV/c range, they are not the
fragments of any hard parton at much higher $p_T$. They are the recombination products
of semi-hard partons at lower $p_T$. Since such semi-hard shower partons are copiously
produced at LHC, treating the detected hadron as a trigger does not select any subset
of events that are characterized by anything special.  It is not like a jet that is
rare at RHIC.  For particles correlated to the trigger on the same side, the shower
partons in the two jets that give rise to the trigger particle must have other sister
shower partons in the same two jets to recombine to form associated particles. Of
course, that can occur, but they are a few among many other recombined products of
an ocean of shower partons with higher yield. Such associated particles merge with
the background and cannot contribute to recognizable peak on the near side. This is
in sharp contrast to the situation at RHIC where such near-side peaks in $\Delta\eta$
and $\Delta\phi$ have been calculated in the recombination model and found to reproduce
the data very well \cite{ch,fbm}.

When a trigger particle is formed in the way that we have described in this paper, the recoil
partons of the two relevant jets are not different from any of the other hard partons that are abundant. Those
recoil partons, if they emerge at all after surviving the quenching effect of the medium
that they traverse, are weaker  than the ambient minijets produced near the surface on
the far side. Hence, we expect no peaks in the away-side $\Delta\phi$ distribution that rise above the background.

In summary, we have shown that when the jet density is high, as is expected at LHC,
the recombination of shower partons arising from adjacent jets can give rise to
hadron production that dominates over parton fragmentation. At constant jet-overlap probability $\Gamma$ the $p/\pi$ ratio increases with $p_T$, but with a reasonable $p_T$ dependence ascribed to $\Gamma(p_T)$ the ratio decreases with $p_T$ from roughly 20 to 5 in the interval $10<p_T<20$ GeV/c. That is still about 2 orders of magnitude higher than the ratio one expects from fragmentation. This prediction offers a striking target for
experimental test of the validity of recombination in an area where fragmentation
has conventionally been applied. Our prediction that there are no associated particles
beyond the uncorrelated background to be found in connection with trigger particles in
the $p_T$ range considered can easily be checked experimentally. Data on these
predictions can provide crucial guidance to the understanding of the hadronization
problem at high $p_T$ that has until now been uncontroversial.

     This work was supported, in part,  by the
U.\ S.\ Department of Energy under Grant No. DE-FG02-96ER40972  and by
the Ministry of Education of China under Grant No. 03113, and by National Natural
Science Foundation of China under Grant No. 10475032.

\end{document}